\documentclass[letterpaper, 10 pt, conference]{ieeeconf}  

\IEEEoverridecommandlockouts                              
\usepackage{graphicx} 
\usepackage{epsfig} 
\usepackage{times} 
\usepackage{amsmath} 
\usepackage{amssymb}  
\usepackage{blindtext}
\usepackage{algpseudocode}
\usepackage{upgreek}
\usepackage{physics}
\usepackage{hyperref}
\usepackage{cleveref}
\usepackage{algorithm}

\DeclareMathOperator*{\argmin}{arg\,min}

\title{\LARGE \bf
	Experiment Design with Gaussian Process Regression with Applications to Chance-Constrained Control}

\author{Sean Anderson, Katie Byl, João P. Hespanha
	\thanks{This material is based upon work supported by the National Science Foundation Graduate Research Fellowship under Grant No. 2139319. Any opinion, findings, and conclusions or recommendations expressed in this material are those of the authors(s) and do not necessarily reflect the views of the National Science Foundation.}
	\thanks{The authors are with Department of Electrical and Computer Engineering, University of California Santa Barbara, Santa Barbara, CA, 93106 USA
		{\tt\small seananderson@ucsb.edu, katiebyl@ece.ucsb.edu, hespanha@ece.ucsb.edu}}%
}

\begin{document}

	\maketitle
	\thispagestyle{empty}
	\pagestyle{empty}

\begin{abstract}
	Learning for control in repeated tasks allows for well-designed experiments to gather the most useful data. We consider the setting in which we use a data-driven controller that does not have access to the true system dynamics. Rather, the controller uses inferred dynamics based on the available information. In order to acquire data that is beneficial for this controller, we present an experimental design approach that leverages the current data to improve expected control performance. We focus on the setting in which inference on the unknown dynamics is performed using Gaussian processes. Gaussian processes not only provide uncertainty quantification but also allow us to leverage structures inherent to Gaussian random variables. Through this structure, we design experiments via gradient descent on the expected control performance with respect to the experiment input. In particular, we focus on a chance-constrained minimum expected time control problem. Numerical demonstrations of our approach indicate our experimental design outperforms relevant benchmarks.
\end{abstract}

\section{Introduction}

Many safety-critical tasks such as driving around a race track or repeated robotic motions in a novel environment can leverage learning-based methods to improve control performance. In the setting where the dynamics of the system are unknown, a natural question is how to use a fixed number of trials to gather the most informative data for improving control performance. 

We motivate our problem with the example of a race car driver learning the nuances of the vehicle dynamics on a new track. If the driver is given test laps before having to race, the driver should choose each test lap carefully so as to improve their expected performance on the race lap. For our problem, we consider a controller that minimizes an objective (e.g. racing) given dynamics inferred from gathered data. In order to collect the most useful data for control, we design an experiment that aims to improve the data-driven controller's expected performance as much as possible. 

In this work we present a novel formulation for experiment design for data-driven control. Section \ref{sec:iter_learn} defines the system dynamics as an unknown discrete-time process with additive noise. We define a data-driven controller that minimizes a control objective given the currently available data. When thinking about experiment design, this naturally leads to a formulation in which we consider how to augment our dataset in order to the improve our control performance as much as possible. We design the experiment to minimize the expected control performance given data available at the time of the experiment.

In Section \ref{sec:gp_reg}, we focus our attention on using Gaussian processes for inference on the system dynamics. In safety-critical settings where we need uncertainty quantification of both epistemic and aleatoric uncertainty of the dynamics, Gaussian processes allow us to derive tractable formulations for both the optimal control and experiment design problems. 

Using this structure, we present an approach to solving the experiment design problem that takes the gradient of the expected control performance with respect to the experiment input. We derive a numerical approximation to the analytic gradient and then use stochastic gradient descent to improve the expected performance.

We consider a prototypical chance-constrained control problem in Section \ref{sec:min_time} in which we try to minimize the expected time to a target set while avoiding an unsafe set to a desired probability level. In particular, we focus on joint chance constraints in time such that the total probability of entering the unsafe set is minimized as in  \cite{ono_chance-constrained_2015}, \cite{paulson_stochastic_2020}. We solve this using a dynamic programming approach and a Lagrangian formulation for the joint-chance constraint. 

Numerical results in Section \ref{sec:num_results} demonstrate the effectiveness of the experiment design formulation for the minimum time problem. We consider the scenario in which we add one well-designed experiment to an existing dataset. We compare the performance of our algorithm against suitable benchmarks by first considering the percentage of experiments that result in feasible controllers in low-data settings. We then also consider the minimum time for controllers that are feasible. These indicate that our experiment design outperforms the benchmarks and highlights potential future work.

Using Gaussian process regression for learning dynamical models for the purposes of safe control has been explored in numerous works including \cite{hewing_cautious_2020, vallon_data-driven_2022}. These papers focus on using Gaussian processes trained on sufficient data to perform safety-critical control. Exactly how to choose that data when given limited time to gather data is not explored. Furthermore, active learning using Gaussian processes has been demonstrated in  \cite{jain_learning_2018,le_receding_2021,buisson-fenet_actively_2020}. Active learning is often used in the dual control context \cite{mesbah_stochastic_2018}; the dual control problem differs from this work because we separate out data gathering from the control trial while a dual control approach would try to simultaneously perform control and estimate unknown quantities. Some work has been done on experiment design for Gaussian processes, mainly with respect to information theoretic objectives and not necessarily for dynamical systems \cite{boukouvalas_approximately_nodate, weaver_computational_2016,hoffmann_mobile_2010,noauthor_near-optimal_nodate, fiorino_gaussian_nodate,houlsby_bayesian_2011}. 

Overall then, our approach differs from previous work in mainly two aspects: first, the experiment design criteria differs from information theoretic approaches by using the expected control performance in order to focus learning in control-relevant regions, and second, our approach to solving the resulting experiment design is novel in how it leverages the structure of Gaussian processes.

\section{Experiment Design for Data-Driven Control} \label{sec:iter_learn}

We consider systems with discrete-time dynamics 
\begin{align}
	x_{t+1} &= f(x_t, u_t) + w_t, \label{eq:gen_dynamics}
\end{align} 
where $x_t \in \mathbb{R}^{n_x}$ denotes the state at time $t$, $u_t \in \mathbb{R}^{n_u}$ denotes the input, and $w_t \sim \mathcal{N}(0,\Sigma_w)$ is independent, identically distributed process noise. We do not know the function $f(\cdot)$ and will use data gathered from previous trials to inform our belief of the values $f(\cdot)$ will take at particular state-input pairs. We consider a control objective given by 
\begin{align}
	G(X,U)
\end{align}
over time horizon $N$ such that the matrix $X \in \mathbb{R}^{n_x \times N}$ is a concatenation of the states, $x_t$, and similarly $U \in \mathbb{R}^{n_u \times N}$. 

While $f(\cdot)$ is not known, we have prior information about it such that we can condition our expected control objective on it. We incorporate available data by letting the set: 
\begin{align}
	\mathcal{D} &= \{x_i, u_i, y_i\}_{i=0}^{M}
	\intertext{consist of $M$ triples satisfying}
	y_i &= f(x_i, u_i) + w_i \label{eq:measure}
\end{align}
with the understanding that the data come from a repeated task in which case (\ref{eq:gen_dynamics}) provides triples that satisfy (\ref{eq:measure}) for $y_i=x_{t+1}, x_i=x_{t}, u_i=u_t$.

After gathering some data, $\mathcal{D}$, we want to conduct a trial that minimizes an expected control objective (e.g. racing a car around a track as fast as possible). In this data-driven setting, our control objective is to minimize a conditional expectation based on the available information and the process noise:
	\begin{align}
		J(\mathcal{D}) &:= \min_{U \in \mathcal{U}} ~\mathbb{E}[G(X,U) \mid \mathcal{D}]  \label{eq:gen_control}
	\end{align}
where $X$ is described by (\ref{eq:gen_dynamics}) such that $X$ is a function of $U$. The expectation is equivalently expressed as $\int G(X,U) p(X \mid U, \mathcal{D}) dX$. 

In order to gather the data that is most useful for the control performance, we construct our experiment design criteria to minimize the post-experiment expected cost of the optimal control problem. To this effect we construct an experiment that will generate a new dataset to augment the existing dataset $\mathcal{D}$ such that if we were to race after the experiment trial, our control performance would be as good as possible:
\begin{align}
	\min_{\bar{U} \in \mathcal{U}} ~\mathbb{E}\big[J\big(\mathcal{D}_{+}(\bar{X},\bar{U})\big) \mid \mathcal{D}\big], \label{eq:gen_exp_des}
\end{align} 
where $\bar{U}$ is the control signal for the new dataset, $\bar{X}$ the corresponding state trajectory from (\ref{eq:gen_dynamics}), $\bar{Y}$ the corresponding measurements, and $\mathcal{D}_{+}(\bar{X}, \bar{U}):=\mathcal{D} \cup (\bar{X},\bar{U},\bar{Y})$. The expectation here is with respect to the possible trajectory, $\bar{X}$, with the current understanding of the dynamics given the data is $\mathcal{D}$. This can be expressed as $\int J\big(\mathcal{D}_{+}(\bar{X},\bar{U})\big) p(\bar{X} \mid \bar{U}, \mathcal{D}) d\bar{X}$.

The experiment design problem then is an offline, bilevel optimization problem that can be written as 
\begin{align}
	\min_{\bar{U} \in \mathcal{U}} ~\mathbb{E}\bigg[ \min_{U \in \mathcal{U}} \mathbb{E} \big[G(X,U) \mid \mathcal{D}_{+}(\bar{X},\bar{U})\big] ~\big{\rvert}~ \mathcal{D}\bigg]
\end{align}
where $X$ and $\bar{X}$ are described by (\ref{eq:gen_dynamics}), conditioned on the respective available information, $\mathcal{D}^+(\bar{X},\bar{U})$ and $\mathcal{D}$.

\section{Gaussian Process Experiment Design} \label{sec:gp_reg}
We now move from the general setting to one in which uncertainty in $f(\cdot)$ is modeled as a Gaussian process.

\subsection{Model inference}
A Gaussian process (GP) is a collection of indexed random variables that are jointly Gaussian, any subset of which is also jointly Gaussian. We assume $f(x,u)$ is a Gaussian process indexed by state-input pairs $z:=(x, u)^T$ with mean, $m(\cdot)$, and covariance function, $k(\cdot,\cdot)$:
\begin{align}
m(z) &= \mathbb{E}[f(z)], \\
k(z_k,z_j) &= \mathbb{E}[(f(z_k) - m(z_k))(f(z_j) - m(z_j))^T].
\end{align}
When comparing two points, $z_k, z_j$, the kernel $k(\cdot,\cdot)$ shapes the covariance of the two random variables. In particular, we use the squared exponential kernel, $k(z_k,z_j)=\sigma_f^2 \exp\big(\frac{-(z_k-z_j)^T(z_k-z_j)}{2\ell^2}\big)$, which is infinitely differentiable \cite{mchutchon_nonlinear_nodate}.

\textit{Corollary 1}: \cite{rasmussen_gaussian_2006}. Assume $f(\cdot)$ is a Gaussian process and the process noise covariance is diagonal such that $\Sigma_w=\text{diag}([\sigma_{w,1}^2,...,\sigma_{w,n_x}^2])$. Then the conditional distribution of the $d$-th entry of $x_{t+1}$ ($d \in \{1,...,n_x\}$) given the dataset $\mathcal{D}$, the state $x_{t}=x$, and the control input $u_t=u$ is a normal distribution with mean and variance:
\begin{subequations} \label{eq:post_gp}
\begin{align} 
		\mu_d(z;\mathcal{D}) &:= m^d(z) + K_{z \mathbf{Z}}^d(K_{\mathbf{Z}\mathbf{Z}}^d + \sigma_{w,d}^2 I)^{-1}(\mathbf{Y}^d- m^d(\mathbf{Z})) \label{eq:post_mean} \\  
		\sigma_d^2(z;\mathcal{D}) &:= K_{zz}^d- K_{z\mathbf{Z}}^d(K_{\mathbf{Z}\mathbf{Z}}^d+\sigma_{w,d}^2I)^{-1}K_{\mathbf{Z}z}^d + \sigma_{w,d}^2 \label{eq:post_var}
\end{align}  
\end{subequations}
where $\mathbf{Z}:=[z_0,...,z_M]$ is the matrix of training indices and $\mathbf{Y}:=[y_0,...,y_M]$ is a matrix of noisy measurements in $\mathcal{D}$. $K$ are Gram matrices that are composed of kernel evaluations: $[K_{\mathbf{Z}\mathbf{Z}}^d]_{lj}=k^d(z_l,z_j),[K_{\mathbf{Z}z}^d]_{j}=k^d(z_j,z),K_{zz}^d=k^d(z,z)$.

\begin{proof}
\begin{subequations}
We consider multi-output GPs with independent outputs and diagonal process noise matrix $\Sigma_w=\text{diag}([\sigma_{w,1}^2,...,\sigma_{w,n_x}^2])$. For each triple $(x_i,u_i,y_i)$ in $\mathcal{D}$, (\ref{eq:measure}) provides a noisy measurement of the GP. We incorporate this in the prior on our observations such that 
\begin{align}
	\mathbf{Y}^d \sim \mathcal{N}(m^d(\mathbf{Z}),K_{\mathbf{Z}\mathbf{Z}}^d+\sigma_{w,d}^2I).
\end{align}
We go on to express the joint distribution of the measurements at observed points, $\mathbf{Y}^d$ corresponding to $\mathbf{Z}$, and at $y^d$ corresponding to a new test point $z$:
\begin{align}
\begin{bmatrix}
	\mathbf{Y}^d \\
	y^d
\end{bmatrix} &\sim \mathcal{N} \Bigg(
\begin{bmatrix}
	m^d(\mathbf{Z}) \\
	m^d(z)
\end{bmatrix},
\begin{bmatrix}
	K_{\mathbf{Z}\mathbf{Z}}^d + \sigma_{w,d}^2 I & K_{\mathbf{Z}z}^d\\
	K_{z\mathbf{Z}}^d & K_{zz}^d
\end{bmatrix} \Bigg)
\end{align}
This joint distribution, $p(\mathbf{Y}^d, y^d \mid z, \mathbf{Z})$, can be conditioned on the dataset $\mathcal{D}$ using $p(y^d \mid \mathbf{Y}^d, z, \mathbf{Z}) = p(\mathbf{Y}^d, y^d \mid z, \mathbf{Z}) / p(\mathbf{Y}^d)$ to obtain the conditional distribution of $y^d$ given $\mathcal{D}$ and $z$ as a normal distribution with mean and variance 
\begin{align}
	\mu_d(z;\mathcal{D}) &= m^d(z) + K_{z \mathbf{Z}}^d(K_{\mathbf{Z}\mathbf{Z}}^d + \sigma_{w,d}^2 I)^{-1}(\mathbf{Y}^d - m^d(\mathbf{Z})) \label{eq:post_mu_der} \\
	\sigma_d^2(z; \mathcal{D}) &= K_{zz}^d - K_{z\mathbf{Z}}^d(K_{\mathbf{Z}\mathbf{Z}}^d + \sigma_{w,d}^2 I)^{-1}K_{\mathbf{Z}z}^d + \sigma_{w,d}^2. \label{eq:post_var_der}
\end{align}
Since $y=f(x,u)+w$, by (\ref{eq:gen_dynamics}) we have that $y=x_{t+1}$ for a given trajectory. We then have that the conditional distribution of $x_{t+1}$, given $\mathcal{D}$ and a particular state and input, is a normal distribution with moments (\ref{eq:post_mu_der},\ref{eq:post_var_der}).

Our GPs are independent such that the predictive equations are given by stacking the individual outputs with diagonal matrix $\Sigma(z;\mathcal{D}):=\text{diag}([\sigma_1^2(z;\mathcal{D}),...,\sigma_{n_x}^2(z;\mathcal{D})])$ and vector $\mu(z;\mathcal{D}):=[\mu_1(z;\mathcal{D}),...,\mu_{n_x}(z;\mathcal{D})]^T$. We refer to the specific index set $(x,u)$ instead of $z$ for clarity going forward.
\end{subequations}
\end{proof}

\subsection{Experiment Design} \label{sec:gp_exp_design}
We develop a gradient descent approach to the experiment design optimization (\ref{eq:gen_exp_des}) as described in Algorithm 1. We derive a gradient estimator that uses Monte Carlo sampling to approximate the expectation in (\ref{eq:gen_exp_des}). The resulting algorithm takes in the available data $\mathcal{D}$, an initial guess for the optimization variable $\bar{U}$, and the initial condition for the experiment $\bar{x}_0$; using these and a Monte Carlo sample from the GP, we generate a possible experiment trajectory. For each sample, we solve the optimal control problem in (\ref{eq:gen_control}) and then compute the gradient of the objective function with respect to $\bar{U}$. We use a batch of samples to perform stochastic gradient descent until a stopping condition is met.

In particular, stochastic gradient descent on the variable $\bar{U}$ requires computation of
\begin{align}
	\pdv{}{\bar{U}} \mathbb{E}[J(\mathcal{D}_{+}(\bar{X},\bar{U})) \mid \mathcal{D}]. \label{eq:gen_grad}
\end{align} 
Defining $h(\bar{X},\bar{U}) := J(\mathcal{D}_+(\bar{X},\bar{U}))$ and denoting by $p(\bar{X} \mid \bar{U})$ the $\bar{U}$-dependent conditional probability density function of $\bar{X}$ given $\mathcal{D}$, we can rewrite (\ref{eq:gen_grad}) as
\begin{align}
	\pdv{}{\bar{U}} \int h(\bar{X},\bar{U}) p(\bar{X} \mid \bar{U}) d\bar{X}.
\end{align}
When both $h(\cdot)$ and $p(\cdot)$ are differentiable with respect to $\bar{U}$, this derivative is given by
\begin{align}
	\int \pdv{h(\bar{X},\bar{U})}{\bar{U}} p(\bar{X} \mid \bar{U})d\bar{X}+ \int h(\bar{X},\bar{U})\pdv{p(\bar{X} \mid \bar{U})}{\bar{U}} d\bar{X}.
\end{align}
While the first term in this expression can be easily computed using Monte Carlo integration, the second is generally more difficult. The following result uses the so-called ``reparameterization trick" \cite{kingma_vae_2022} to obviate this difficulty, allowing us to directly use Monte Carlo integration to compute (\ref{eq:gen_grad}).

\begin{subequations}
\textit{Theorem 1:} Suppose the GP regressor (\ref{eq:post_gp}) has a kernel and mean function both differentiable with respect to the index. Furthermore, assume that the value of the objective function is differentiable with respect to $\bar{U}$. Then we can construct an estimate of (\ref{eq:gen_grad}) using L samples:
\begin{align}
	\frac{1}{L} \sum_{l=1}^{L} \pdv{c(v_0^l,v_1^l,...,v_{N-1}^l,\bar{U})}{\bar{U}}. \label{eq:grad_est}
\end{align}
Here $v_i, ~i \in \{0,...,N-1\}$, are standard normal random variables and 
\begin{align}
	\begin{split}
		c(&v_0^l,v_1^l,...,v_{N-1}^l,\bar{U}) := \\
		&J\big(\mathcal{D}_{+}([\bar{x}_0,\mu(\bar{x}_0,\bar{u}_0)+\Sigma(\bar{x}_0,\bar{u}_0)^{\frac{1}{2}}v_0^l,\\
		&\mu(\bar{x}_1,\bar{u}_1)+\Sigma(\bar{x}_1,\bar{u}_1)^{\frac{1}{2}}v_1^l,..., \\
		&\mu(\bar{x}_{N-1},\bar{u}_{N-1})+\Sigma(\bar{x}_{N-1},\bar{u}_{N-1})^{\frac{1}{2}}v_{N-1}^l], \bar{U})\big).
	\end{split} \label{eq:c_def}
\end{align}
$\bar{x}_t$ is sampled from the recursively-defined distribution $\mathcal{N}\big(\mu(\bar{x}_{t-1}\bar{u}_{t-1}), \Sigma(\bar{x}_{t-1},\bar{u}_{t-1})\big)$ starting at the initial state $x_0$ with $\mu(\cdot)$ and $\Sigma(\cdot)$ given by (\ref{eq:post_gp}).
In our numerical examples the gradient in (\ref{eq:grad_est}) of (\ref{eq:c_def}) is computed by automatic differentiation.
\end{subequations}

\begin{proof}
\begin{subequations}
Starting from the joint-density describing the forward simulation of the dynamics given $\mathcal{D}$, we express the expected value in (\ref{eq:gen_grad}) as
\begin{align}
	\begin{split}
		\mathbb{E}\big[J\big(\mathcal{D}_{+}(&\bar{X},\bar{U})\big) \mid \mathcal{D}\big] = \\
		& \int J\big(\mathcal{D}_{+}(\bar{X},\bar{U})\big) p(\bar{X} \mid \bar{x}_0, \bar{U}, \mathcal{D}) d\bar{X}, \label{eq:gp_criteria}
	\end{split}
\end{align}
where the density function describes the distribution of the trajectory given the input sequence $\bar{U}$, the currently available data $\mathcal{D}$, and the process noise for each time step. Using Bayes' rule for probability density functions, the joint distribution of the states $p(\bar{X} \mid \bar{x}_0, \bar{U}, \mathcal{D})$ can be recursively expanded as
\begin{align}
	\begin{split}
	p(\bar{X} \mid \bar{x}_0, \bar{U}, \mathcal{D})& \\
	=p(\bar{x}_N&,...,\bar{x}_2 \mid  \bar{x}_0, \bar{x}_1, \bar{U}, \mathcal{D})p(\bar{x}_1 \mid \bar{x}_0, \bar{U}, \mathcal{D}). \\
	=p(\bar{x}_N&,...,\bar{x}_3 \mid\bar{x}_0, \bar{x}_1, \bar{x}_2, \bar{U}, \mathcal{D}) \times \\
	p(\bar{x}_2& \mid \bar{x}_0, \bar{x}_1, \bar{U}, \mathcal{D}) p(\bar{x}_1 \mid \bar{x}_0, \bar{U}, \mathcal{D}). \\
	= \Pi_{t=1}^N& p(\bar{x}_t \mid  \bar{x}_0, ....,\bar{x}_{t-1}, \bar{U}, \mathcal{D}).
	\end{split}
\end{align}
As a consequence of (\ref{eq:gen_dynamics}), our dynamics are Markovian such that we have
\begin{align}
	\begin{split}
	p(\bar{x}_t \mid  \bar{x}_0, ....,\bar{x}_{t-1}, \bar{U}, \mathcal{D}) = 
	p(\bar{x}_t \mid  \bar{x}_{t-1}, \bar{u}_{t-1}, \mathcal{D}).
	\end{split}
\end{align}
From Corollary 1, we know $p(\bar{x}_t \mid  \bar{x}_{t-1}, \bar{u}_{t-1}, \mathcal{D})$ is described by (\ref{eq:post_gp}) for a particular state-input pair. Using the reparameterization trick \cite{kingma_vae_2022}, Gaussian random variables can be expressed in terms of a standard normal, $v$, where in our case
\begin{align} 
	\bar{x}_{t+1}=\mu(\bar{x}_t,\bar{u}_t;\mathcal{D}) +\Sigma(\bar{x}_t,\bar{u}_t;\mathcal{D})^{\frac{1}{2}}v, \label{eq:reparam_x}
\end{align}
and $\Sigma(\bar{x}_t,\bar{u}_t;\mathcal{D})^{\frac{1}{2}}$ is a diagonal matrix of the standard deviation of each output dimension $d$ since from (\ref{eq:post_gp}) the outputs are independent. Given $\bar{x}_0$, we apply this change of variables recursively starting at $\bar{x}_1$:
\begin{align}
	\mathbb{E}\big[J\big(\mathcal{D}_{+}(&\bar{X},\bar{U})\big) \mid \mathcal{D}\big] = \\
	\begin{split}
	\int...\int J\big(&\mathcal{D}([\bar{x}_0,\mu(\bar{x}_0,\bar{u}_0)+\Sigma(\bar{x}_0,\bar{u}_0)^{\frac{1}{2}}v_0,\bar{x}_2,...,\bar{x}_N], \bar{U})\big) \\
	&\times p(v_0) \Pi_{t=2}^{N} p(\bar{x}_t \mid  \bar{x}_{t-1}, \bar{u}_{t-1}, \mathcal{D}) dv_0d\bar{x}_2...d\bar{x}_N \label{eq:reparam_int}
	\end{split}
\end{align}
since $d\bar{x}_{t+1} = \Sigma(\bar{x}_t,\bar{u}_t)^{\frac{1}{2}} dv_t$, and for a scalar normal random variable $v_t$ the change of variable leads to $p(\mu+\sigma v_t | \bar{x}_t,\bar{u}_t) = \frac{1}{\sqrt{2\pi}\sigma}\exp(-\frac{1}{2}v^2) \sigma = p(v_t)$; we can extend this to our problem since the outputs of (\ref{eq:post_gp}) are independent such that the joint distribution can be considered a product of the individual density functions. We continue the recursion, where $x_t$ is recursively computed based on (\ref{eq:reparam_x}), and for clarity define
\begin{align}
	\begin{split}
		c(&v_0,v_1,...,v_{N-1},\bar{U}) := \\
		&J\big(\mathcal{D}_{+}([\bar{x}_0,\mu(\bar{x}_0,\bar{u}_0)+\Sigma(\bar{x}_0,\bar{u}_0)^{\frac{1}{2}}v_0,\\
		&\mu(\bar{x}_1,\bar{u}_1)+\Sigma(\bar{x}_1,\bar{u}_1)^{\frac{1}{2}}v_1,..., \\
		&\mu(\bar{x}_{N-1},\bar{u}_{N-1})+\Sigma(\bar{x}_{N-1},\bar{u}_{N-1})^{\frac{1}{2}}v_{N-1}], \bar{U})\big).
	\end{split}
\end{align}
We express our resulting integral more concisely now as
\begin{align}
	\int ...\int c(v_0,v_1,...,v_{N-1},\bar{U}) \Pi_{t=0}^{N-1} p(v_t) dv_0...dv_{N-1},
\end{align}
which we can approximate numerically as
\begin{align}
	\frac{1}{L} \sum_{l=1}^L c(v_0^l,v_1^l,...,v_{N-1}^l,\bar{U}).
\end{align}
Under the assumption that $J(\cdot)$ and the GP moments (\ref{eq:post_gp}) are differentiable with respect to $\bar{U}$, the Monte Carlo integrator is now differentiable with respect to $\bar{U}$:
\begin{align}
	\frac{1}{L} \sum_{l=1}^L \pdv{}{\bar{U}} c(v_0^l,v_1^l,...,v_{N-1}^l,\bar{U}).
\end{align}
Note that while the differentiability of $J(\cdot)$ depends on the particular objective function, the differentiability for the GP moments is met when using a differentiable kernel and mean function.
\end{subequations}
\end{proof}

\begin{algorithm}
	\caption{\strut Experiment Design}\label{alg:exp_des_algo}
	\label{alg:exp_design}
	\hspace*{\algorithmicindent} \textbf{Input} $\bar{U}~ (\text{initial}), \mathcal{D}, \bar{x}_0$ \par
	\hspace*{\algorithmicindent} \textbf{Output} $\bar{U}$
	\begin{algorithmic}
		\Function{SampleGrad}{$\bar{U},V,\mathcal{D},x_0$} 
		\State $(\bar{X},\bar{U}, \bar{Y}) \gets$ \text{forward sim. from} $(\bar{U},V,\mathcal{D},\bar{x}_0)$
		\State $\mathcal{D}_{+} \gets \mathcal{D} \cup (\bar{X},\bar{U}, \bar{Y})$
		\State $\mu(z;\mathcal{D}_{+}), \Sigma(z;\mathcal{D}_{+}) \gets$ \text{update GP with $\mathcal{D}_{+}$}
		\State $U^*(\bar{U},V) \gets$  \text{Solve control with $\mu(z;\mathcal{D}_{+}), \Sigma(z;\mathcal{D}_{+})$}
		\State $\pdv{J}{\bar{U}}  \gets$  \text{Compute gradient at $U^*(\bar{U},V)$}
		\EndFunction
		
		\While{not converged}
		\For{i=1:batch size}
		\State $V \gets$ \text{$[v_0,v_1,...,v_{N-1}]$ from standard normal}
		\State $\pdv{J(V)}{\bar{U}} \gets SAMPLEGRAD(\bar{U}, V, \mathcal{D}, \bar{x}_0)$
		\EndFor
		\State $\bar{U} \gets$ \text{gradient step from batch, project onto $\mathcal{U}$}
		\EndWhile
	\end{algorithmic}
\end{algorithm}

\begin{figure}[th]
	\begin{center}
		\includegraphics[width=8.4cm]{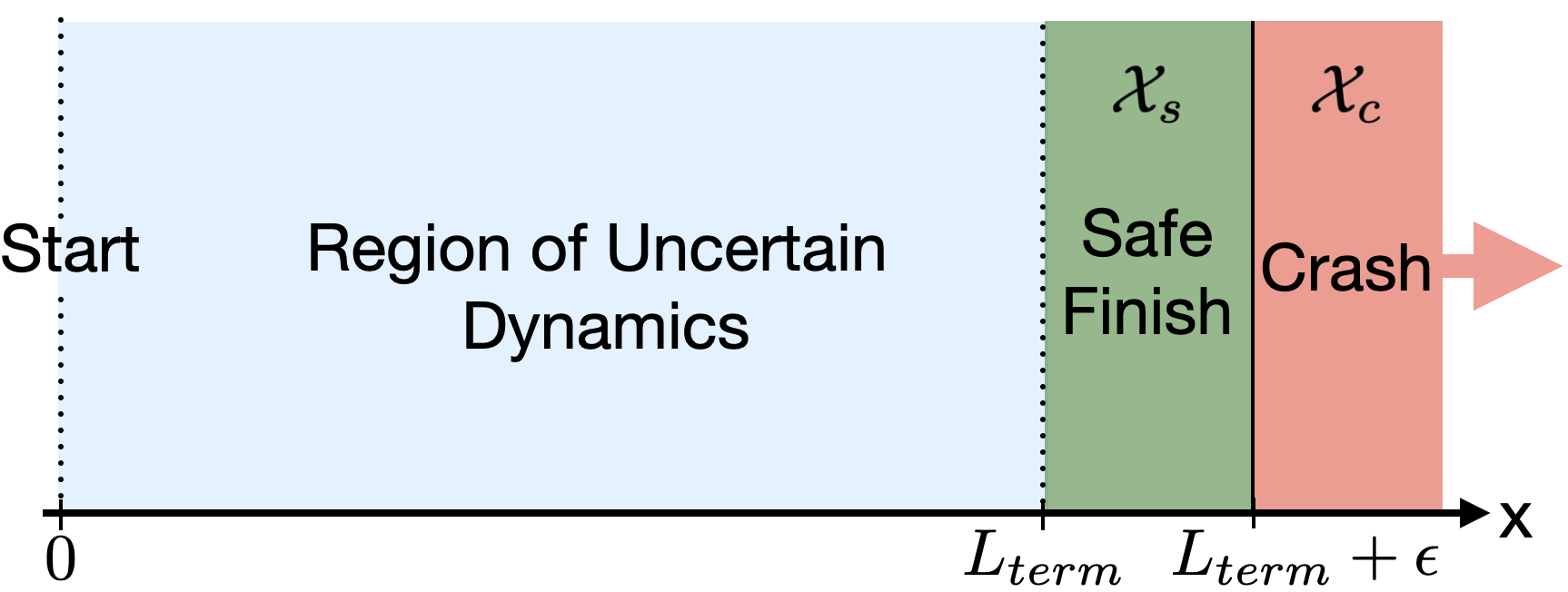}    
		\caption{Starting from the origin, the objective is to reach the green target set, $\mathcal{X}_s$, in minimum time. Overshooting the target set results in a crash in $\mathcal{X}_c$. The dynamics of the system are unknown and need to be learned.} 
		\label{fig:state_space}
	\end{center}
\end{figure}
	
\section{Minimum expected time} \label{sec:min_time}

To demonstrate the approach laid out above, we consider a prototypical stochastic control problem with a chance constraint. We start at the origin and want to reach a safe finishing set, $\mathcal{X}_{s}$, while avoiding an unsafe set, $\mathcal{X}_{c}$, that also leads to termination. Collectively, we refer to these two sets as $\mathcal{X}_{term}$. In a one-dimensional setting we visualize this in Figure \ref{fig:state_space}. Formally this corresponds to the following optimization:

\begin{subequations}  \label{eq:opt_control}
	\begin{align}
		\min_{U} ~& \mathbb{E} [T \mid \mathcal{D}]  \\ 
		s.t. &~\mathbb{E}\bigg[\sum_{t=0}^{T} I(x_t) \mid x_0, \mathcal{D} \bigg] \leq \Delta, \label{eq:risk_bound} \\
		& u_t  \in \mathcal{U}, \\
		& x_0 = 0,
	\end{align} 
\end{subequations}
where $x_t$ is described by (\ref{eq:gen_dynamics}), $\Delta$ is our tolerance for failure and the constraint (\ref{eq:risk_bound}) is a conservative bound on the probability of failure, $P(x_T \in \mathcal{X}_{c}) \leq \Delta$ \cite{ono_chance-constrained_2015}. Here $I(x)$ indicates a state in the unsafe set $\mathcal{X}_{c}$, such that
\begin{align}
	I(x) &= \begin{cases}
		1, &x \in \mathcal{X}_{c}, \\
		0,  &o/w.
	\end{cases} 
\end{align}
We define $T$, which is a random variable, as the first time for which the state enters $\mathcal{X}_{term}$, unless the state never reaches either finished set, in which case $T=N$. Once we enter $\mathcal{X}_{term}$, the state does not change. We express this as
\begin{align}
	T:=\min ~\{t: x_t \in \mathcal{X}_{term}\} \cup \{N\}. 
\end{align} 

\textit{Corollary 2}: 
\begin{subequations}
	\cite{ono_chance-constrained_2015}. Suppose there exists a solution to the Bellman equation 
	\begin{align}
			\begin{split}
					&J_t^{\lambda^*}(x_t) = \\
					&\begin{cases}
							\min_{u} ~1 + \mathbb{E}[J_{t+1}^{\lambda^*}(x_{t+1}) \mid x_t=x, \mathcal{D}], & x_t \notin   \mathcal{X}_{s} \cup \mathcal{X}_{c},\\
							0, &x_t \in \mathcal{X}_{s}, \\
							\lambda^*, &x_t \in \mathcal{X}_{c},
						\end{cases}
				\end{split} \label{eq:bell_cases}
	\end{align}
	with terminal cost-to-go given by
	\begin{align}
		J_N^{\lambda^*}(x_N) = 
		\begin{cases}
			\lambda^*, & x_N \notin \mathcal{X}_{s}, \\
			0, & o/w,
		\end{cases} \label{eq:bell_term}
	\end{align}
	for which trajectories under the policy
	\begin{align}
		\begin{split}
		u_t = \pi(x_t) := 
		&\argmin_{u} ~1 + \mathbb{E}[J_{t+1}^{\lambda^*}(x_{t+1}) \mid x_t=x, \mathcal{D}] \label{eq:bellman_result}
		\end{split}
	\end{align}
	satisfy
	\begin{align}
		\lambda^* \bigg(\mathbb{E}\bigg[\sum_{t=0}^{T} I(x_t) \mid x_0, \mathcal{D} \bigg] - \Delta\bigg)=0,
	\end{align}
	then (\ref{eq:bellman_result}) solves the optimization (\ref{eq:opt_control}).
\end{subequations}

\begin{proof}
\begin{subequations}
For a dynamic programming (DP) problem with stage costs $g_t(x)$ and the risk bound (\ref{eq:risk_bound}), we can form a Lagrangian \cite{ono_chance-constrained_2015}:
\begin{align}
	\begin{split}
			\mathcal{L}(x,&u,\lambda, x_0, \mathcal{D}) = \\
			&\mathbb{E}\bigg[\sum_{t=0}^N g_t(x_t) \mid \mathcal{D}\bigg]
			+ \lambda \bigg(\mathbb{E}\bigg[\sum_{t=0}^{N} I_t(x_t) \mid x_0, \mathcal{D} \bigg] - \Delta\bigg)
		\end{split}
\end{align}
such that the primal problem is given by solving the Bellman equation
\begin{align}
	\begin{split}
	J_t^{\lambda}&(x_t) = \\
	&\min_{u} ~g_t(x_t) + \lambda I_t(x_t) + \mathbb{E}[J_{t+1}^{\lambda} (x_{t+1}) \mid x_t=x, \mathcal{D}]
	\end{split}
\end{align}
 for a fixed $\lambda$ and terminal cost 
	\begin{align}
		J_N^{\lambda}(x_N) = g_N(x_N) + \lambda I_N(x_N).
	\end{align}
The choice and optimality of $\lambda$ follows from \cite{ono_chance-constrained_2015} Section 2 and 4.2, so we provide a sketch of the optimality conditions as they pertain to finding $\lambda^*$. We know the dual problem is concave irrespective of the form of the primal. Since the dual is concave, $\lambda^*$ maximizes $q(\lambda)$ for $\lambda \geq 0$ if and only if there exists a subgradient $d$ in the subdifferential $\partial q$ such that $d^T(\lambda - \lambda^*) \leq 0 ~\forall \lambda \geq 0$. For $\lambda^*=0$, the primal problem is feasible, and for $\lambda^* > 0$, it is necessary and sufficient for $0 \in \partial q(\lambda)$.

Our minimum time problem (\ref{eq:opt_control}) is in discrete-time such that for a DP formulation our stage cost counts unit time for states not finished. By construction, once we enter a finished set we remain there for all future time. We model this using dynamics with $x_t \in \mathcal{X}=\mathbb{R}^{n_x} \cup \{done, bad\}$:
\begin{align}
	\begin{split}
		x_{t+1} := 
		&\begin{cases}
			f(x_t,u_t) + w_t, & x_t \in \mathbb{R}^{n_x}, \\
			done, & x_t\in \mathcal{X}_s \cup \{done, bad\}, \\
			bad, & x_t \in \mathcal{X}_c,
		\end{cases} \label{eq:long_dyn}
	\end{split}
\end{align}
such that \textit{bad} corresponds to reaching $\mathcal{X}_c$ and \textit{done} corresponds to having reached the safe terminal set $\mathcal{X}_s$ or $\textit{bad}$ previously. We then define our stage cost $g(x_t)$ to be 
\begin{align}
	g_t(x_t) :=
	\begin{cases}
		1, & x_t \in \mathbb{R}^{n_x}, \\ 
		0, & o/w,
	\end{cases} \label{eq:long_stage}
\end{align}
and the cost-to-go as
\begin{align}
	\begin{split}
		J_t(x_t&)= 
		\begin{cases}
			\min_u ~\mathbb{E}[J_{t+1}^{\lambda} (x_{t+1}) \mid x_t=x, \mathcal{D}], &x_t \in \mathbb{R}^{n_x}, \\
			0, &x_t=done, \\
			\lambda, &x_t=bad. \\
		\end{cases} \label{eq:long_j}
	\end{split}
\end{align}
This model satisfies the Lagrangian approach above such that we directly apply it. Additionally, the model based on (\ref{eq:long_dyn}) to \ref{eq:long_j}) can be more compactly expressed since for $x \in \mathcal{X}_{s}$, the cost-to-go is zero; for $x \in \mathcal{X}_c$, the cost-to-go is $\lambda$; and for $x \notin \mathcal{X}_s \cup \mathcal{X}_c$, the cost-to-go is 
\begin{align}
	J_t^\lambda(x_t) = \min_u ~1 + \mathbb{E}[J_{t+1}^{\lambda} (x_{t+1}) \mid x_t=x, \mathcal{D}]
\end{align} 
with 
\begin{align}
	J_N^\lambda(x_N) = 
	\begin{cases}
		\lambda, & x_N \notin \mathcal{X}_{s}, \\
		0, & o/w,
	\end{cases}
\end{align}
giving us the compact expression in (\ref{eq:bell_cases},\ref{eq:bell_term}). The expected cost-to-go is with respect to the conditional distribution (\ref{eq:post_gp}) from Corollary 1:
\begin{align}
	\begin{split}
		\mathbb{E}[&J_{t+1} (x_{t+1}) \mid x_t, \mathcal{D}] =\\
		& \int J_{t+1}(x_{t+1})  p(x_{t+1} \mid x_t=x, u), \mathcal{D}) dx.
	\end{split}
\end{align} 

\end{subequations}
\end{proof}

In order to find $\lambda^*$, as in \cite{ono_chance-constrained_2015}, we update $\lambda$ based on a root-finding algorithm on the subgradient of the dual function $q(\lambda, x_0, \mathcal{D}) := \min_{u} ~\mathcal{L}(x,u,\lambda, x_0, \mathcal{D})$:
\begin{align}
	\pdv{q(\lambda, x_0, \mathcal{D})}{\lambda} = \mathbb{E}\bigg[\sum_{t=0}^{N} I_t(x_t) \mid x_0, \mathcal{D} \bigg] - \Delta.
\end{align}
When the subgradient is sufficiently close to zero, we have $\lambda^*$.

\section{Numerical Results} \label{sec:num_results}
We present numerical simulations for the minimum expected time problem. To show the benefit of optimal experiment design, we compare the performance improvement of conducting an experiment using our method versus other intuitive methods. 

\subsection{Process model}
\begin{figure}[thpb]
	\begin{center}
		\includegraphics[width=8.4cm]{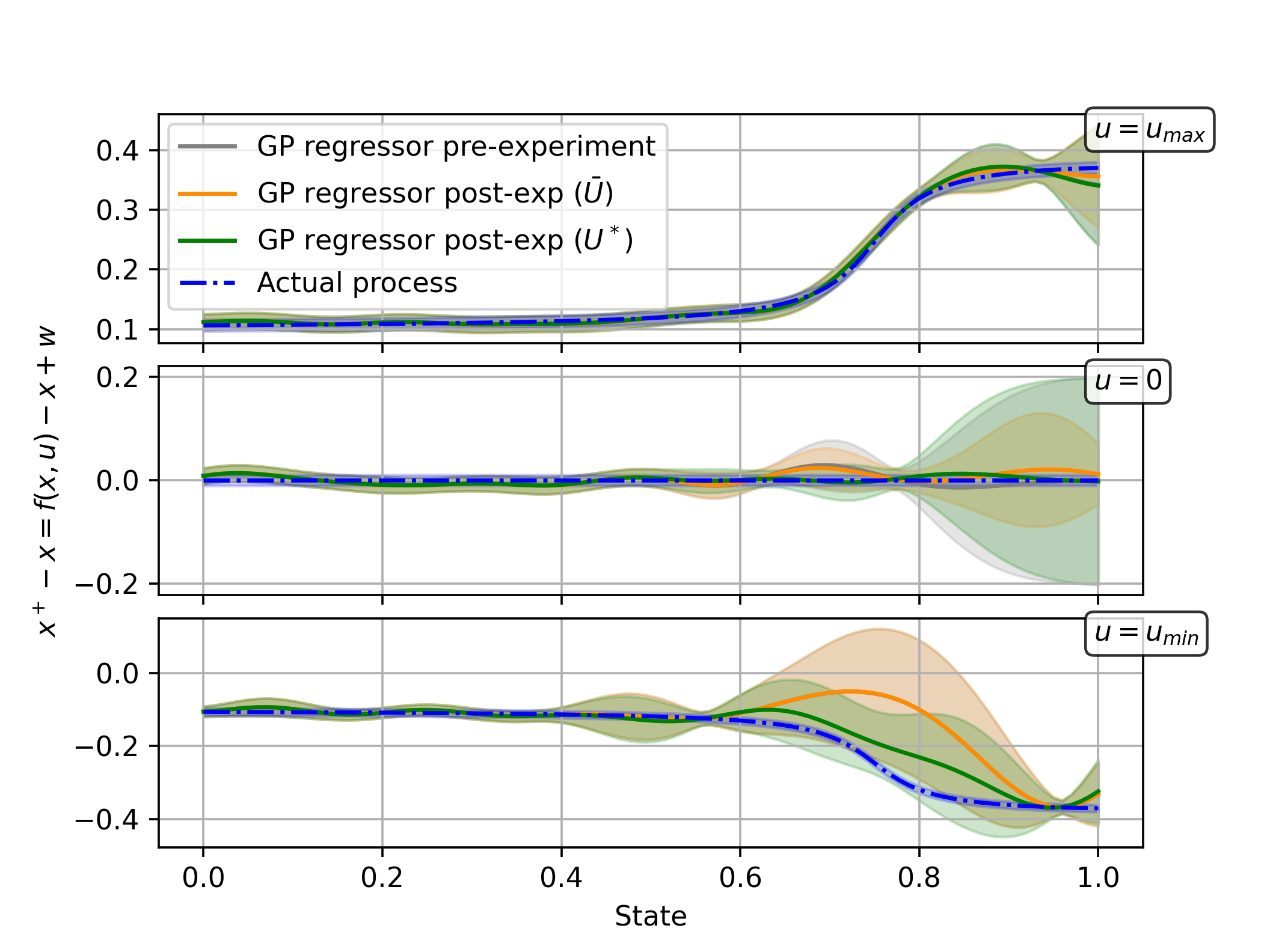}    
		\caption{We visualize our GP approximation of the process mean and variance by taking slices of the input at three different levels, $u_{max}=0.1,0,u_{min}=-0.1$, and sweep the state. In blue, we observe the actual process with relatively small 1-$\sigma$ bounds from the process noise. The pre-experiment GP's dataset has 25 random trials, while the post-experiment GPs' datasets contain the original 25 plus the respective experiment trial. For $u_{max}$, we observe the GP models have relatively low epistemic uncertainty (variance), which is reasonable given the random walks have positive mean. Our experiment input, $\bar{U}$ reduces the uncertainty for zero-input significantly relative to the pre-experiment or $U^*$. The closed-loop, $U^*(x)$ is left off here for legibility. While $U^*$ appears to generate a model with less uncertainty for $u_{min}$, this is less relevant since the optimal control input does not need to be negative in this setup.}
		\label{fig:moments}
	\end{center}
\end{figure}

Consider a system modeled by (\ref{eq:gen_dynamics}) with the initially unknown scalar function
\begin{align}
	f(x,u) &= x + \big(1+3\tilde{I}_{[0.75,1.4]}(x)\big)u 
\end{align}
where $\tilde{I}(\cdot)$ is the indicator function approximated by product of arctangents with finite slope. This can be visualized in the dark blue trace in Figure \ref{fig:moments}. We set the process noise to $\Sigma_w=0.01^2$.

For the control problem in (\ref{eq:opt_control}), we start at $x_0=0$ and try to finish in the set $\mathcal{X}_s=[L,L+\epsilon]=[1,1.05]$ (Figure \ref{fig:state_space}). Overshooting $\mathcal{X}_s$ enters the unsafe set $\mathcal{X}_c=(1.05,\infty)$. Our tolerance for failure in (\ref{eq:risk_bound}) is set to $\Delta=25\%$. For both the control problem and experiment design, we restrict the control input to $[-0.1, 0.1]$. The max allowable time, $N$, is 15 time steps for both the control and experiment design problems.

\subsection{Benchmarks}
We start with a dataset, $\mathcal{D}$, consisting of a fixed number of random walk trials with positive mean. We then create $\mathcal{D}_{+}$ by augmenting $\mathcal{D}$ with an additional trajectory using our method (Algorithm 1) referred to as $\bar{U}$ or one of the benchmarks described below. Based on $\mathcal{D}_{+}$ we compute the new optimal control (\ref{eq:opt_control}) and evaluate the control performance. In our comparisons, we consider the bounding performance value given perfect knowledge of the process, $f(\cdot)$. In terms of expected time and risk, this is the ideal performance and is referred to as ``perfect info" in plots. For the benchmarks, we use the following alternatives to our experiment design:
\begin{itemize}
	\item \textit{random}: add another random walk trajectory to the set $\mathcal{D}$.
	\item $U^*$: compute an open-loop control based on the optimal control (\ref{eq:opt_control}) given $\mathcal{D}$ for the most-likely state sequence sampled from our current GP. This open-loop control also serves as the initial value for Algorithm 1.
	\item $U^*(x)$: use the optimal state feedback policy from (\ref{eq:opt_control}) given $\mathcal{D}$. While we consider open-loop input sequences for our experiment design, including the relative performance of a closed-loop policy shows the value of feedback in this stochastic setting.
\end{itemize} 
Each experiment trajectory is run for the duration of the experiment time and does not stop upon reaching either of the target sets. This can lead to shorter trajectories if the state reaches the finished set in less than 15 steps for the optimal control benchmarks. However, for the small dataset scenarios that we explore in the coming section, the optimal control tends to use the full-time since it cannot guarantee constraint satisfaction. 

\subsection{Computations}
For efficient programming, we used JAX \cite{noauthor_jax_2023} and GPJAX \cite{pinder_gpjax_2022} in Python such that the stochastic DP can be just-in-time compiled and we can simultaneously use automatic differentiation in (\ref{eq:grad_est}). We used 1000 stochastic gradient descent steps instead of waiting until convergence, with a batch size of 80 at each step on a dual core NVIDIA GeForce RTX 3080 GPU.

\subsection{Discussion}
\begin{figure}[tb]
	\begin{center}
		\includegraphics[width=8.4cm]{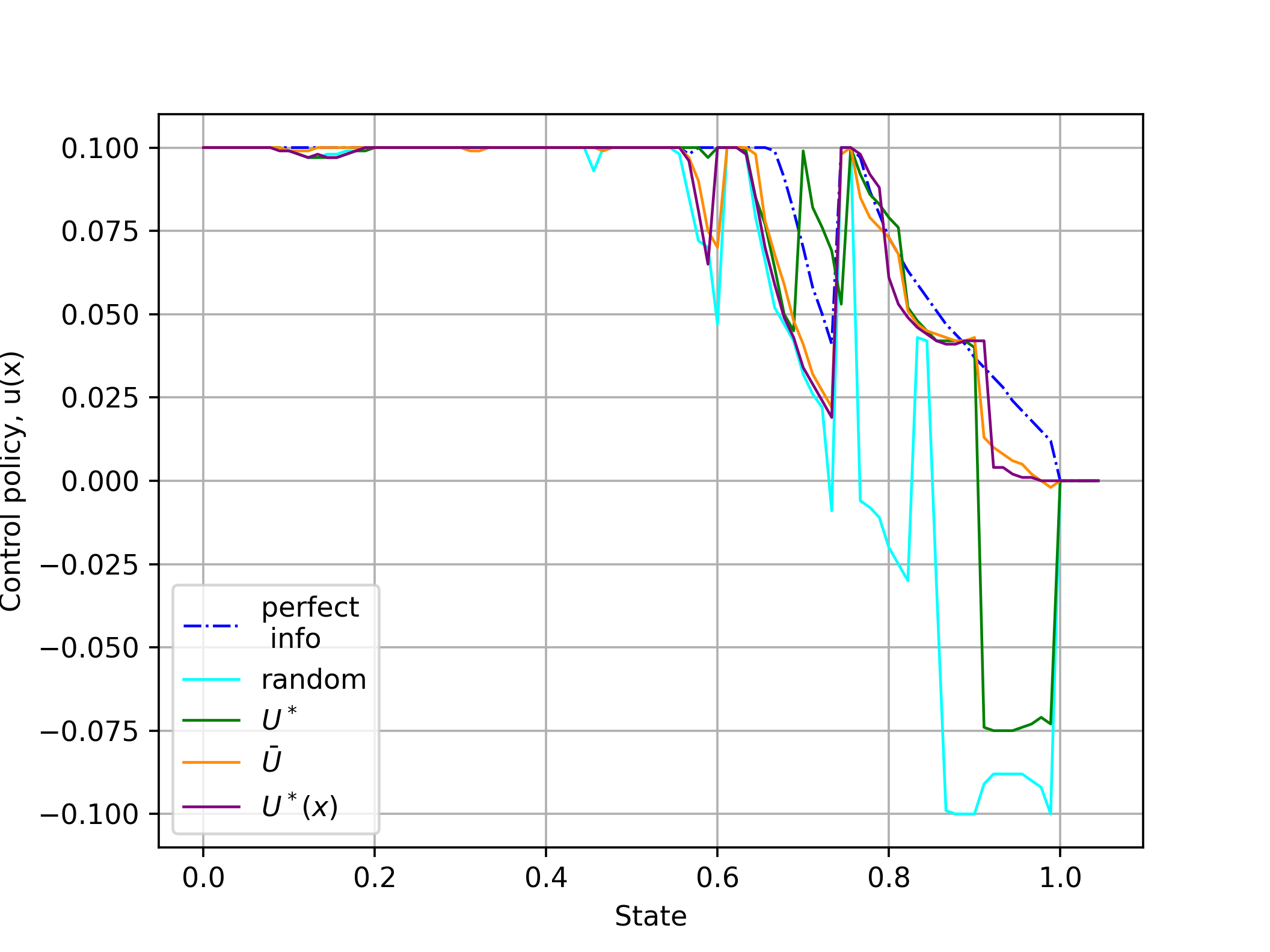}    
		\caption{We illustrate the optimal policies post-experiment. Since we have slightly different policies depending on the experiment outcome, we illustrate the policy closest to the median for our experiment design and the corresponding outcome for the benchmarks. We observe that the policy given perfect information is to apply the max input, 0.1, until the state reaches 0.6, then decrease until around 0.75, and then decrease the input to zero after just before 0.8. In this example, our experiment design $\bar{U}$ tends to approach the optimal without exceeding it. In some areas the benchmarks are closer to the perfect information policy but they also exceed it, which leads to riskier behavior.}
		\label{fig:policy}
	\end{center}
\end{figure}

Figure \ref{fig:moments} helps build intuition about the performance of our method relative to the benchmarks. While we will later discuss the behavior over various initial dataset sizes, this figure depicts the uncertainty of three different GP models for the particular case of 25 initial random trials: the pre-experiment GP has access to 25 trials of data, the GP corresponding to $\bar{U}$ contains the same 25 trials plus our well-designed experiment, and the GP for $U^*$ contains the 25 initial trials plus the trial from applying $U^*$. We leave off the closed-loop comparison to avoid cluttering the plot. We observe that our method reduces epistemic uncertainty mainly for zero-input near the finishing set. Figure \ref{fig:policy} illustrates that given the post-experiment datasets, our experiment design leads to a control policy that approaches the optimal policy given perfect information. The optimal control with perfect information decreases to zero near the target set, which indicates the observed reduction of epistemic uncertainty at zero in Figure \ref{fig:moments} by our method is useful for improving control performance. Unlike the benchmarks, our method does not exceed the optimal policy given perfect information, doing which leads to higher risk. This qualitative analysis indicates that our experiment design, $\bar{U}$, leads to learning in areas that are important for control performance. 

\begin{figure}[tb]
	\begin{center}
		\includegraphics[width=8.4cm]{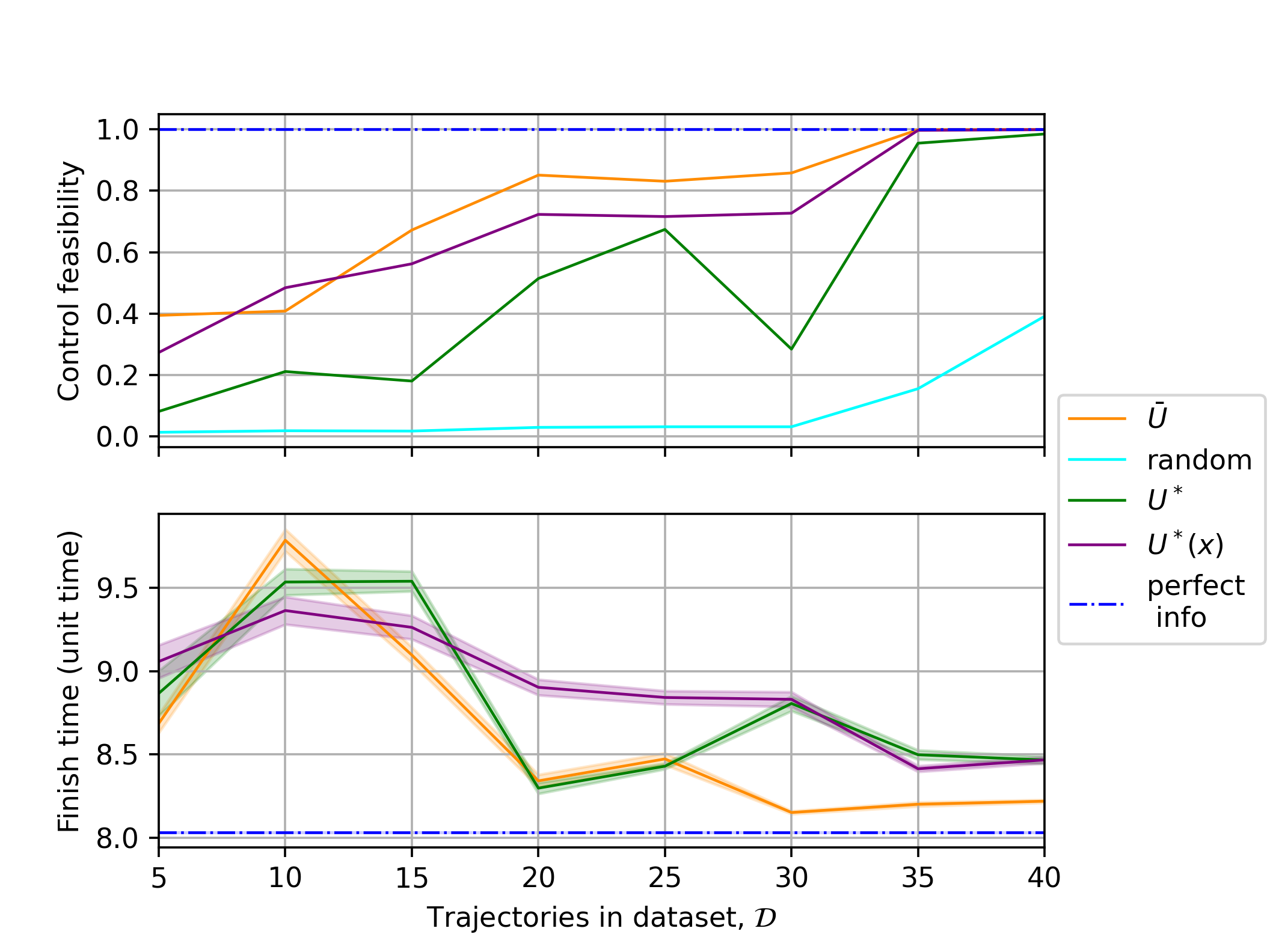}    
		\caption{We visualize the benefit of adding a designed experiment to a dataset of random trials (e.g. 5 random trials plus one experiment) and test the control performance with this augmented dataset. Control feasibility indicates the fraction of controllers that satisfy the chance constraint over 1000 experiment outcomes. Of the controllers that satisfy the constraint, we can compute the expected finish time and 95 percent confidence intervals. Our experiment design, $\bar{U}$, outperforms all other benchmarks for control feasibility except for $U^*(x)$ for the dataset of size 10. Finishing time performance becomes more relevant as the feasibility metric nears 100 percent whereupon we see our time decreasing significantly relative to the others. Due to low control feasibility, we omit the finish time for the random case.} 
		\label{fig:trends}
	\end{center}
\end{figure}

We quantify the experiment design performance using two metrics: the fraction of controllers (\ref{eq:opt_control}) that can satisfy the chance constraint (\ref{eq:risk_bound}), and from the subset of controllers that are feasible, the expected time to finish. In Figure \ref{fig:trends}, we show the trends for the two metrics across dataset size. In this setting, 40 random trials are drawn and each dataset of size $n$ is a superset of dataset size $m$ for $n > m$ (i.e. the 10 trial dataset includes the 5 trial dataset plus 5 additional trials). In the first subplot of Figure \ref{fig:trends} we observe the fraction of controllers that satisfy our risk tolerance level, $\Delta=25\%$, where the data is from 1000 possible experiment outcomes for each experiment design method. This shows that our method in general generates a controller that satisfies the constraint more than any of the benchmarks. While we significantly outperform the open-loop optimal control, we also do better than the closed-loop optimal control, which has the advantage over all other methods of having access to state feedback. One takeaway from including the closed-loop optimal control is that closed-loop experiment design methods can increase performance significantly. Lastly, while we might expect monotonic improvement with respect to dataset size across all methods, this is not the case because we are dealing with a particular set of 40 random trials.

The second subplot in Figure \ref{fig:trends} illustrates the expected time for feasible controllers. For the smaller datasets, the significance of this metric is much secondary to the control feasibility because most controllers do not satisfy the chance constraint. We omit the finish time for ``random" since the control feasibility is very low for all datasets. It is worth noting that besides the dataset of size 10, our method is comparable in expected time or better than the benchmarks. For larger datasets, we observe that as the control feasibility approaches 100 percent, our expected time decreases significantly relative to the others as desired.
	

\addtolength{\textheight}{-3cm}   


\section{Conclusion}
We present a general approach to experiment design that improves data-driven control performance as much as possible  when the dynamics are initially unknown. We focus on using Gaussian processes for inference and derive a tractable formulation for our experiment design optimization. We consider a minimum expected time problem with chance constraints and numerically demonstrate that our experiment design method outperforms suitable benchmarks. While the minimum expected time control problem was shown in a particular setting, this demonstrates an instance of our broader experiment design approach. 

This work scratches the surface for experiment design for stochastic optimal control problems where the goal of the design is to improve the control performance. Finding efficient relaxations that reduce the need to recompute an optimal solution for each sample of the gradient can improve the complexity of this approach significantly. For Gaussian process regressors, efficient augmentation of the data matrix will enable faster computations such as via rank-n updates for Cholesky decompositions. Additionally, preliminary testing indicates that designing an experiment policy will allow for greater performance than the current open-loop methodology. Finally, adding inequality constraints to the experiment design problem is critical for real-world applications.
	
%
	

\bibliographystyle{IEEEtran}
\bibliography{IEEEabrv,bibliography}

%
%
%
%

\end{document}